\documentclass[twocolumn,12pt,cites,graphicx]{article}
\newcommand{\half} {\frac{1}{2}}
\newcommand{\dv} { {\rm d} }
\newcommand {\scharge} {\left|\sigma\right|}
\newcommand {\cs} {c_{\rm salt} }
\newcommand{\yy} {\zeta}

\usepackage{epsfig}

\title{Bending Moduli of Charged Membranes Immersed in
         Polyelectrolyte Solutions\\}

\author{Adi Shafir\thanks{E-mail:shafira@post.tau.ac.il} and
    David Andelman\thanks{E-mail:andelman@post.tau.ac.il}\\
    School of Physics and Astronomy   \\
    Raymond and Beverly Sackler Faculty of Exact Sciences \\
    Tel Aviv University, Ramat Aviv, Tel Aviv 69978, Israel \\
     }

\date{Received 1st September 2006, corrections 28th December
    \\DOI: 10.1039/}

\begin{document}

\maketitle
\renewcommand{\thefootnote}{\fnsymbol{footnote}}

\noindent We study the contribution of polyelectrolytes in solution
to the bending moduli of charged membranes. Using the Helfrich free
energy, and within the mean-field theory, we calculate the
dependence of the bending moduli on the electrostatics and
short-range interactions between the membrane and the
polyelectrolyte chains. The most significant effect is seen for
strong short-range interactions and low amounts of added salt where
a substantial increase in the bending moduli of order $1\, k_BT$ is
obtained. From short-range repulsive membranes, the polyelectrolyte
contribution to the bending moduli is small, of order $0.1\,k_BT$ up
to at most $1\,k_BT$. For weak short-range attraction, the increase
in membrane rigidity is smaller and of less significance. It may
even become negative for large enough amounts of added salt. Our
numerical results are obtained by solving the adsorption problem in
spherical and cylindrical geometries. In some cases the bending moduli
are shown to follow simple scaling laws.

\section{Introduction}
\label{intro}

The study of interactions between charged and flexible membranes and
polyelectrolytes (PEs) in solution has generated a lot of interest
in recent years, partly motivated by the importance of such
interactions in biology.  Understanding the interaction between
charged macromolecules such as DNA, RNA and various proteins and
biological cell membranes (modeled as charged and flexible
interfaces) sheds light on many important cellular processes.
Besides its biological significance, the adsorption of PEs onto
charged membranes raises interesting questions about the interplay
between short-range  and electrostatic (long range) interactions in
these multi-component charged systems. Recent works include
numerical and analytical calculations
~\cite{itamar1,itamar2,itamar3,us1,us2,wang1,wiegel,muthu,joanny,varoqui1,varoqui2}
and scaling arguments~\cite{itamar1,us1,us2,borisov,dobrynin}.

In our previous work~\cite{usphase}, three regimes have been found
for polyelectrolyte adsorption. (i)~When the {\it short-range
interaction} between the membrane and the PE chains is repulsive,
the surface charge is low and the ionic strength of the solution is
high, the polymers deplete from the charged membrane. (ii)~For
higher surface charges, or lower ionic strength, the
polyelectrolytes adsorb on the charged membrane and screen the
surface charges. (iii)~When the short-range interactions between the
membrane and the PE chains are attractive, the PE chains adsorb on
the membrane, and the adsorbed layer carries a higher charge than
that of the bare membrane. In this situation the polyelectrolytes
over compensate the bare surface --- a phenomenon of practical
importance in the build-up of multilayers of alternating cation and
anion polyelectrolytes~\cite{decher_book}.

The flexibility of fluid surfaces and membranes has been studied in
various cases, and it depends on the lipid composition, tail length
and molecular tilt. In the case of charged and flexible membranes
immersed in a pure ionic solution (no macromolecules)
~\cite{bensimon,fodnin,winter,winter2,KK,lekker,mitchnin,fodmitchnin,andelman,deVries,scheutjens,cstuart},
the electrostatic contribution was found to increase the membrane
rigidity, and the addition of salt to decrease it. In a different
set of studies, the adsorption of neutral polymers on membranes has
been investigated
theoretically~\cite{degennes91,marques,joannycle,skau,chlipo,sunglee,netzlipo}
and in experiments~\cite{ligoure1,ligoure2,gommper,maugey}. It was
found that the addition of polymers reduces the membrane rigidity.
In all those cases, the contribution of the solution to the bending
rigidity was found to be of order $0.1-1\,k_BT$, which is low in
comparison to the intrinsic monolayer bending rigidity of
approximately $10\,k_BT$. We note that in special cases, by adding a
co-surfactant (alcohol), the membrane bending rigidity can be
brought down to values of roughly $1\,k_BT$ \cite{RouxSafinya}.

In the present work, we study the combined system of charged chains
interacting with oppositely charged and flexible membranes. Our
study is similar in spirit to that done for DNA-lipid systems
\cite{harries1,harries2}, where the DNA was modeled as rigid and
charged rod. The main difference is that our charged macromolecules
are flexible. The above mentioned three regimes dictate a different
contribution to the membrane rigidity and stability. For short-range
repulsive membranes, the membrane rigidifies due to its charges
while its rigidity decreases with the increase of the
polyelectrolyte charge. For weak short-range interactions, the
contribution to the membrane rigidity decreases and may become
slightly negative, while for strong short-range attraction the
membrane becomes rigid again. In most cases, the magnitude of the
contribution to the bending rigidity is of the same order of
magnitude as that of neutral polymers or pure salt solutions, namely
$0.1\,k_BT$ up to 1\,$k_BT$. However, we show in this paper that for
strong enough short-range attraction a more significant
contribution, of order $k_BT$ or higher can be obtained.

\section{Mean Field Equations} \label{MFEq}

\subsection{Free-Energy Formulation}
\label{fe}

Consider a bulk aqueous solution containing polyelectrolyte (PE)
chains, along with their counter ions and  added  salt. A curved and
charged membrane is placed at the origin $\left|{\bf r}\right|=0$.
In order to extract the membrane elastic moduli, we take the
membrane shape to be  either spherical or cylindrical with radius
$R$ . The free energy for the combined system has been formulated
before~\cite{itamar1}, and can be written as a sum of four
contributions $F_{\rm tot}=F_{\rm pol}+F_{\rm ions}+F_{\rm
el}+F_{\rm int}$.

The polymer free energy, $F_{\rm pol}$, is:

\begin{eqnarray}
  F_{\rm pol} & = & \int_V\dv {\bf r}
  \Bigg[\frac{a^2}{6}\left|\nabla\phi\right|^2+
  \Sigma\left(\phi\right)-\Sigma\left(\phi_b\right)+
\nonumber  \\ & & \qquad\qquad
    \mu\left(\phi^2-\phi_b^2\right)\Bigg],
 \label{Fpol}
\end{eqnarray}
where $a$ is the monomer size, $\phi_b^2$ is the monomer bulk
concentration, $\phi^2$ is the local monomer
concentration. The free energy is given in terms of $k_BT$, and the
integration is carried out over the entire volume outside the
sphere/cylinder, $\left|{\bf r}\right|>R$. The first term in
Eq.~(\ref{Fpol}) accounts for the chain elasticity. The second and
third terms account for the excluded volume interactions in the
solution and in the bulk, respectively:

\begin{eqnarray}
  \Sigma \left(\phi\right)a^3 &= & \left(1-a^3\phi^2\right)
    \log\left(1-a^3\phi^2\right)- \nonumber \\ & &
    \half\left(v-a^3\right)
    \left(1-a^3\phi^2\right)\phi^2
 \label{lambdaEq}\\
    \mu & = &     \log\left(1-a^3\phi_b^2\right)
    + \nonumber \\ & & \half\left(v a^{-3}+1\right)-
    \left(v-a^3\right)\phi_b^2
 \label{MuEq}
\end{eqnarray}
where $v$ is the excluded volume coefficient. For $\phi^2\ll
a^{-3}$ the above expression can be expanded to the well known
$\half v\left (\phi^2-\phi_b^2\right)^2$, which is commonly used
for the excluded volume interaction.

The small ion contribution to the entropy is:

\begin{equation}
  F_{\rm ions}= \sum_{i=\pm}\int_V\dv {\bf r}
    \left[ c_i\log \frac{c_i(x)}{c_b^i}-c_i(x)+c^i_b\right]
 \label{Fion}
\end{equation}
where $c_\pm$ and $c_b^\pm$ are the local and bulk concentrations of
the $\pm$ small ions, respectively. The third contribution to the free
energy, the electrostatic free-energy, is:

\begin{eqnarray}
  F_{\rm el} & = & \int_V\dv {\bf r}
  \left(c_+-c_-+f\phi^2\right)\yy - \nonumber \\
   & & ~ \frac{1}{8\pi l_B} \int_V\dv{\bf r}
  \left|\nabla\yy\right|^2+
    \int_{\left|{\bf r}\right|=R}\dv {\bf A} \sigma\yy
 \label{Fel}
\end{eqnarray}
where $\yy\equiv e\psi/k_BT$ is the renormalized electrostatic
potential, $f$ is the fraction of charged monomers and $l_B\equiv
e^2/\varepsilon k_BT$ is the Bjerrum length. The first integral
accounts for the interactions between the positive ions, negative
small ions and the monomer charges with the electrostatic potential.
The second integral accounts for the self-energy of the
electrostatic field and the third to the interaction between the
surface charge and the electrostatic potential. Note that the third
integral is taken only over the charged body surface, $\left|{\bf
r}\right|=R$.

The last part of the free energy is the short-range interaction of
the PE chains with the membrane:

\begin{equation}
  F_{\rm int} =
    - \frac{a^2}{6d}\int_{\left|{\bf r}\right|=R}\dv{\bf A}\,
   \phi_s^2
 \label{Fint}
\end{equation}
where $\phi_s$\,$\equiv$\,$\phi\left(\left|{\bf r}\right|=R\right)$
is the monomer concentration on the surface. This term stands for a
general short-range interaction, where the interaction strength is
determined by the phenomenological constant $d^{-1}$, which has
dimensions of inverse length.

Minimization of the total free energy $F_{\rm tot}=F_{\rm
pol}+F_{\rm ions}+F_{\rm el}+F_{\rm int}$ yields the following
mean-field equations, expressed in terms of a dimensionless variable
$\eta\equiv\phi/\phi_b$:

\begin{eqnarray}
   \nabla^2\yy & = & \lambda_D^{-2}\sinh\yy +k_m^2\left({\rm e}^\yy
   -\eta^2\right)- \nonumber \\ & & \qquad \qquad
    4\pi l_B\sigma\delta\left(\left|{\bf r}\right|-R\right)
 \label{pbed}\\
   \frac{a^2}{6}\nabla^2\eta & = &
    \Lambda\left(\eta\right)\eta-f\yy\eta \nonumber \\
    &  & \qquad \qquad
    -\frac{a^2}{6d}\delta\left(\left|{\bf r}\right|-R\right)\eta
 \label{edwards}
\end{eqnarray}
where $\lambda_D=\left(8\pi l_B \cs\right)^{-1/2}$  is the
Debye-H\"{u}ckel length scale for the screening of the electrostatic
potential in the presence of added salt and $k_m^{-1}=\left(4\pi l_B
\phi_b^2 f\right)^{-1/2}$ is the corresponding length for the
potential decay due to counterions. Note that the actual decay of
the electrostatic potential is determined by a combination of salt,
counterions, and polymer screening effects. The excluded volume
interaction is taken into account using the function:

\begin{eqnarray}
  \Lambda\left(\eta\right) & = & \log(1-\phi_b^2a^3)-\log(1-\phi_b^2a^3\eta^2)+ \nonumber \\
         & & \quad \left(v-a^3\right)\phi_b^2\left(\eta^2-1\right)
 \label{fullxv}
\end{eqnarray}
which represents the full excluded volume interaction.

The solution of Eqs.~(\ref{pbed}) and (\ref{edwards}) requires four
boundary conditions. Two of the boundary conditions are taken far
from the membrane, where the monomer concentration and electrostatic
potential retrieve their bulk values $\eta\left(\left|{\bf
r}\right|\rightarrow\infty\right)\rightarrow 1$ and
$\yy\left(\left|{\bf r}\right|\rightarrow\infty\right)\rightarrow
0$. The other two boundary conditions account for the interaction
with the charged membrane, and can be obtained by integrating
Eqs~(\ref{pbed}) and (\ref{edwards}) from $\left|{\bf r}\right|=R$ to a
small distance from the membrane, yielding:

\begin{eqnarray}
  \hat{\bf n}\cdot\left.\nabla\yy\right|_{\left|{\bf r}\right|=R} & = &
    -4\pi l_B\sigma
 \label{elbc} \\
  \hat{\bf n}\cdot\left.\nabla\eta\right|_{\left|{\bf r}\right|=R} & = &
    -d^{-1}\eta_s
 \label{chembc}
\end{eqnarray}
Equation~(\ref{elbc}) is the usual electrostatic boundary condition
for a given surface charge density, while Eq.~(\ref{chembc}) is the
Cahn - de Gennes boundary condition~\cite{degennes}, which is often
used for calculating polymer profiles
~\cite{joannycle,skau,marques}.

For large $R$, the total free-energy can be expanded around its flat
surface value in the following way~\cite{helfrich}:

\begin{eqnarray}
  F_{\rm tot}& = & \int_{\left|{\bf r}\right|=R} \dv{\bf A}
    \Bigg[f_0  +
    \half\delta\kappa\left(c_1+c_2-\frac{2}{R_0}\right)^2+\nonumber \\
    & & \qquad \qquad \quad \delta\kappa_G c_1c_2\Bigg]
 \label{helfrich}
\end{eqnarray}
where $F_{\rm tot}$ is the total free energy (in units of $k_BT$),
$f_0$ is the free energy per unit area of a solution in contact with
a planar surface $\left(R\to\infty\right)$
that has the same system parameters, and
$c_1,\,c_2$ are the radii of curvature for the membrane. For
spherical surfaces the radii of curvature are
$c_1$\,$=$\,$c_2$\,$=$\,$1/R$, while for a cylindrical surface
$c_1$\,$=$\,$1/R,\,c_2$\,$=$\,$0$. The parameters $\delta\kappa$ and
$\delta\kappa_G$ are the contributions of the PE (and salt) solution
to the mean and Gaussian curvature moduli of the membrane,
respectively, in units of $k_BT$. Namely, $\kappa=\kappa^0+\delta
\kappa$  and $\kappa_G=\kappa^0_G+\delta \kappa_G$ include the
intrinsic values as well as the contributions coming from the
solution. Throughout this paper we will consider only the changes in
the elastic moduli $\kappa, \kappa_G$ with respect to their bare
values. An increase in the mean curvature modulus $\delta\kappa$
increases the membrane rigidity, while an increase in the Gaussian
modulus $\delta\kappa_G$ makes saddle points on the membrane more
favorable.

The parameter $R_0$ is the radius of spontaneous curvature for the
membrane, which is dependent on the exact chemical composition of
the membrane. For a positive $R_0$, the membrane bends towards the
external solution, while for negative $R_0$ it bends away from the
solution. In the case of a bilayer membrane, where the same solution
is in contact with both leaflets of the membrane, the spontaneous
curvature vanishes due to symmetry. In this paper, we do not
calculate the spontaneous curvature but rather focus on the changes
to the bending moduli $\delta\kappa$ and $\delta\kappa_G$.

\subsection{Numerical Procedure}
\label{numpro}

Equations~(\ref{pbed})-(\ref{chembc}) are solved numerically for the
cases of a charged sphere and a charged cylinder. The numerical
procedure follows the relaxation scheme~\cite{nr}, as was described
in previous publications~\cite{us1,us2}. For each solution, we
calculate the total free energy per unit area for the cases of a
spherical membrane $f_S$ and a cylindrical membrane $f_C$. The
contributions of the polyelectrolyte solution to the mean and
Gaussian curvatures are then calculated by expanding $f_C,\,f_S$ to
second order in $1/R$:

\begin{eqnarray}
  f_S & = & f_0+\frac{A_S}{R}+\frac{B_S}{R^2}
 \label{Fsphere} \\
  f_C & = & f_0+\frac{A_C}{R}+\frac{B_C}{R^2}
 \label{FCyl}
\end{eqnarray}
The contributions of the solution to the curvature moduli are given
by:

\begin{eqnarray}
  \delta\kappa & = & 2B_C
 \label{kappaEq}\\
  \delta\kappa_G & = & B_S-4 B_C.
 \label{kappaGEq}
\end{eqnarray}

A surface is stable under long wave bending fluctuations only when
$\kappa$\,$>$\,$0$, and against spontaneous vesiculation
(topological change) when  $2\kappa+\kappa_G>0$~\cite{andelman}. In
the following we show that the contribution of the PE solution to
the stability of a charged surface depends on the amount of added
salt, and has a non-monotonic dependence on the short-range
interactions between the membrane and the polyelectrolyte.

\section{Results } \label{numerical}

We find large contributions, of order $1\,k_BT$, to the surface
bending moduli for the case of strong short-range attractive
surfaces. For weaker short-range interactions and for repulsive
surfaces, the contribution is smaller, of order $0.1-1\,k_BT$. We
discuss first the strong repulsive and strong attractive surface
limits, and then turn to the intermediate case, where the
contribution to the bending rigidity is less significant. We present
analogies and scaling calculations to explain the different regimes.

\subsection{Strong Short-Range Attractive Membranes}
\label{SRAtt}

In previous publications~\cite{us2,wang1}, the adsorbed amount of
PEs as well as the layer width were studied in detail for
short-range attractive surfaces. The adsorbed PEs charge was shown
to exceed the bare surface charge significantly for strong
short-range interactions, and to exceed it mildly for weak
short-range interactions. The width of the adsorbed layer, on the
other hand, depends on the shorter of the two
adsorption length scales: (i) $d$, the length scale for short-range
attraction from Eq.~(\ref{chembc}), and (ii) $\xi\equiv
a/\sqrt{v\phi_b^2}$, the Edwards correlation length for neutral
polymer adsorption. In our model, we use an almost theta solvent
$0<v\phi_b^2\ll 1$, so we assume that the shorter length scale is
always $d\ll\xi$. The increase in short-range attractive interactions, in
this case,  decreases the adsorbed layer width~\cite{us1,us2}. The
combined charge of the PE-membrane complex is, therefore, opposite
to the initial surface charge, and its magnitude may be much higher
than the initial membrane charge. This charge is distributed within
a layer of width $d$ close to the surface.

Our focus in this section is on the low salt case. In Fig.~1 we show
the dependence of $\kappa$ and $\kappa_G$ on the short-range
interaction parameter $d^{-1}$ for low salt conditions. For strong
short-range attraction, namely $d<2$\AA, the calculated
$\delta\kappa$ is positive and increases strongly, reaching values
of several $k_BT$. The $\delta\kappa_G$ is negative, and shows a
stronger increase with $d^{-1}$ than the corresponding
$\delta\kappa$. The contribution to the vesiculation stability
$2\delta\kappa+\delta\kappa_G$ in this case can be seen to be
negative for high $d^{-1}$, leading to destabilization of the
surface. For lower (but positive) $d^{-1}$, the contribution to
$2\delta\kappa+\delta\kappa_G$ is positive, and thus enhances the
membrane stability.

These results can be explained by the following argument. When the
adsorbed layer width is smaller than the electrostatic screening
length, the membrane-PE complex can be viewed as a renormalized
charged surface, containing both the bare surface charges and the
adsorbed PE charges, in a weak ionic solution. The renormalized
surface interacts with the ionic solution in the same manner as a
bare
membrane~\cite{bensimon,fodnin,winter,winter2,KK,lekker,mitchnin,fodmitchnin,andelman}.
The strong increase in the surface charge and the lack of small ion
screening (low salt) cause the surface fluctuations to be strongly
unfavorable, making $\delta\kappa$ positive. This allows substantial
increase in the magnitude of $\delta\kappa,\,\delta\kappa_G$,
amounting to several $k_BT$, which is a significant contribution to
the membrane curvature moduli, as can be seen in Fig.~1. We note
that as the renormalized surface charge increases, the value of
$\delta\kappa$ should approach the low salt limit for charged
surfaces in pure ionic solutions (no PE) $\delta\kappa\to \lambda_D/
(2\pi l_B)\simeq 7\,k_BT$~\cite{andelman} (See Fig.~1). However,
this limit is still higher than our numerical results for the
PE-membrane complex.

The crossover between positive and negative values of
$2\delta\kappa+\delta\kappa_G$ can also be explained by analogy to
ionic solutions. There are two regimes for
$2\delta\kappa+\delta\kappa_G$ in pure ionic
solutions~\cite{andelman}, depending on the surface charge. For
weakly charged surfaces $\cs\gg l_B\scharge^2$ (or
$\lambda_D\ll\lambda_{\rm GC}$ where $\lambda_{\rm
GC}\equiv\left(2\pi l_B\scharge\right)^{-1}$ is the Gouy-Chapman
length),  $\delta\kappa$ as well as $2\delta\kappa+\delta\kappa_G$
are positive~\cite{winter,KK}, while for highly charged surfaces
$\cs\ll l_B\scharge^2$ we get $\delta\kappa>0$ and
$2\delta\kappa+\delta\kappa_G<0$~\cite{lekker,mitchnin}. In our case
of polyelectrolyte solutions, for low enough attractive short-range
interactions the renormalized surface charge is low, and
$2\delta\kappa+\delta\kappa_G>0$. For stronger short-range
interactions, the renormalized surface charge increases and
$2\delta\kappa+\delta\kappa_G<0$.

The results in Fig.~1 were presented in the low added salt case,
$d\ll \lambda_D$. When more salt is added into the solution the
electrostatic interactions between the polymers become weaker. In
case of a high amount of added salt, $d\gg \lambda_D$, the charges
of the adsorbed polymers are screened over smaller length scales
than the adsorbed layer width, and the analogy to a renormalized
charged surface breaks down. We further discuss this case in
Sec.~\ref{intermed}.

\subsection{Short-Range Repulsive Membranes}
\label{SRRep}

Without the PE adsorption, the charges on the membrane increase the
membrane rigidity, $\kappa$, due to the long-range repulsion between
them. For strongly repulsive surfaces, the adsorption of PE chains
to the membrane screens its surface charges and causes $\kappa$ to
decrease. Note that the PE charges do not fully compensate the
membrane bare charges like they do in the strongly attractive
regime, leading to the difference in the corresponding system
behaviors.

There are two regimes for the case of short-range repulsive
membranes: (i) the low-salt regime where the surface charges are
mainly balanced by the adsorbed monomer charges, and (ii) the
high-salt regime where the salt ions screen the surface charges and
the monomers deplete. The bending moduli in the latter regime are
similar to those of a strong ionic solution with no added PE, as
were discussed in detail in other
papers~\cite{bensimon,fodnin,winter,KK,andelman}.

In Ref.~\cite{us1} we addressed the PE adsorption close to a {\it
flat} charged surface and showed that the screening length of the
electrostatic potential depends strongly on the amount of added
salt. For low salt conditions, the surface charges are screened
mainly by the adsorbed monomer charges. In this case the free-energy
per unit area scales like $f_{\rm ads}\sim
\scharge^{5/3}f^{-1/3}a^{2/3}l_B^{2/3}$ and the screening length
scales like  $D\sim a^{2/3}/(fl_B\scharge)^{1/3}$~\cite{us1}. For
high amounts of added salt, the screening is mainly done by small
ions and the PE chains deplete. The transition from depletion to
adsorption regimes occurs when the screening length due to monomer
adsorption becomes similar to the one for small ion adsorption,
i.e., when $\lambda_D\sim D$.

The flat surface results can now be easily extended to curved
surfaces. In the low-salt regime, monomers are adsorbed to the
membrane, and the length scale for the adsorption is $D$. We expect
the free energy of the curved membrane (per unit area) to scale like
$f_{\rm tot}=f_{\rm ads}h\left(D/R\right)$. Expanding $f_{\rm tot}$
to second order in $R^{-1}$ and comparing to Eq.~(\ref{helfrich})
shows that both curvature moduli scale like:

\begin{equation}
  \delta\kappa,\,\delta\kappa_G\sim f_{\rm ads} D^2\sim \frac{\scharge
  a^2}{f},
 \label{kScRSLS}
\end{equation}
which is of order $0.1-1\,k_BT$ for physiological range of system
parameters.

In Fig.~2 we present the values of $\delta\kappa$ and
$\delta\kappa_G$ as a function of the surface charge $\scharge$, in
the case of strongly repulsive membranes $d^{-1}=-20$\,\AA$^{-1}$
and monomer size of $a=10$\,\AA. For low amounts of added salt we
find a scaling relation
$\delta\kappa,\,\delta\kappa_G\sim\scharge^{\beta}$ with
$\beta\simeq 1.2$. Note that the numerically calculated exponent
$\beta \simeq 1.2$ is  slightly larger than $\beta=1$ derived in
Eq.~(\ref{kScRSLS}).  This discrepancy seems to occurs because the
high monomer size used here makes the full excluded volume
interaction substantial. For comparison, we plot the corresponding
$\delta\kappa,\,\delta\kappa_G$ for the case of a pure ionic
solution, namely when no polyelectrolytes are added to the solution.
In this case the magnitudes of both $\delta\kappa,\,\delta\kappa_G$
are very high. This shows that the addition of polyelectrolytes into
low salt solutions can cause a strong reduction of the bending
moduli, in the order of several $k_BT$ as compared with the pure
salt solution. This reduction can be explained by the fact that the
polyelectrolytes replace the salt ions in screening the surface
charges, thus allowing greater membrane flexibility.

In the high salt regime, the membrane interacts only with the small
ions, and the polymer chains are depleted. The free energy in this
case is similar to that of a weakly charged surface in an ionic
solution~\cite{andelman}. Namely, the screening length is
$\lambda_D$ and the free energy per unit area for the case of a flat
surface is $f_{\rm dep}\sim\scharge^2 l_B \lambda_D$.  The curvature
moduli are the same as in a pure ionic
solution~\cite{bensimon,fodnin,winter,KK,andelman}, where:

\begin{equation}
  \delta\kappa,\,\delta\kappa_G\sim f_{\rm dep}\lambda_D^2 \sim
    \scharge^2l_B\lambda_D^3.
 \label{kScRSHS}
\end{equation}
Note that the depletion condition $\lambda_D\,<\,D$ implies that the
bending moduli for the high salt (depletion) case are always lower
than for low salt (adsorption) case, despite the depletion of the
polymers. Both $\delta\kappa$ and $\delta\kappa_G$ in this case are
of order $0.01-0.1\,k_BT$ for physiological range of system
parameters, and scale like $\scharge^2$.

For both low and high salt regimes, the contribution to the Gaussian
bending modulus $\delta\kappa_G$ is negative, resulting from the
electrostatic repulsion between the membrane
constituents~\cite{winter2}.  The contribution of the PE solution to
the vesiculation stability $2\delta\kappa+\delta\kappa_G$ is always
positive, even for very low salt concentrations, in contrast to the
pure ionic solution results~\cite{lekker,mitchnin,fodmitchnin}. The
low-salt regime of the ionic solutions, in which the contribution to
$2\delta\kappa+\delta\kappa_G$ is negative, is replaced here by the
adsorption regime, in which this contribution is still positive.

\subsection{Weak Short-Range Interacting Membranes}
\label{intermed}

In Fig.~3 we present an enlargement of Fig.~1 for surfaces having
only a weak short-range interaction with the PE in solution,
$d^{-1}\sim 0$. As can be seen, the magnitudes of both
$\delta\kappa$ and $\delta\kappa_G$ decrease substantially for low
$\left|d^{-1}\right|$, and may become negative for high values of
added salt. The decrease in the magnitude of the bending moduli can
be attributed to the strong screening of the surface charges by the
adsorbed PEs, which makes $\delta\kappa$ smaller. We note that the
contribution to $\delta\kappa$ and $\delta\kappa_G$ is negligible in
comparison to the membrane intrinsic bending moduli, and probably
cannot be observed experimentally. This regime is presented only in
order to complete the $d^{-1}$ dependence picture.

For low and positive $d^{-1}$ and high amounts of added salt, we
find a negative contribution to the mean curvature modulus due to
the polymer adsorption $\delta\kappa<0$, as well as in
$2\delta\kappa+\delta\kappa_G$. This surprising result can be
explained by analogy to the neutral polymer case. In past
publications~\cite{marques,joannycle,skau} it was shown that neutral
polymer solutions have negative $\delta\kappa$, positive
$\delta\kappa_G$ and negative $2\delta\kappa+\delta\kappa_G$. We
find here similar results.

The analogy to neutral polymer adsorption is important and can be
understood in the following way. The high amount of added salt
screens both the surface and the PE charges, so their effective
interaction becomes short-ranged leading the way to an almost
neutral polymer behavior. The increase in $\delta\kappa_G$ derives
from the effective attraction between membrane constituents, which
results from their short-range attraction to the PE chains. The
analogy to neutral polymers requires the screening length for the
electrostatic interactions to be much smaller than the layer width
$\lambda_D< d$. This is satisfied in high salt and low short-range
attraction conditions. For stronger short-range interactions, the
layer width decreases, and the electrostatic interactions between
monomers becomes important. In this case, the analogy to neutral
polymers breaks down, and $\delta\kappa,\,
2\delta\kappa+\delta\kappa_G$ start to increase back. For higher
$d^{-1}$, both $\delta\kappa$ and $2\delta\kappa+\delta\kappa_G$
become positive again, marking the crossover to the strong
attraction regime described in Sec.~\ref{SRAtt} above. This can be
seen in Fig.~3, where for high amounts of added salt, the increase
in $\delta\kappa$ with $d^{-1}$ indeed begins when $\lambda_D\simeq
d$, as expected from the above analysis. It is also important to
note, that the magnitude of both $\delta\kappa$ and $\delta\kappa_G$
 are very small, of order $0.01\,k_BT$, and are negligible in
comparison to the intrinsic bending moduli of a membrane, of order
$10\,k_BT$. The decrease shown here cannot account for a decrease in
membrane rigidity that was recently found for DNA
adsorption~\cite{frantescu}.

\section{Conclusions} \label{conclusion2}

In this paper we show that the interaction between charged and
flexible polymers (PE) and oppositely charged and flexible membrane
depends both on their electrostatic and short-range interactions. A
non-monotonic dependence of the curvature moduli is obtained as
function  of the short-range interaction between the membrane and
the PE chains. We find a significant contribution to the bending
moduli, of order of several $k_BT$, in the case of strong
short-range attraction between the PE chains and the surface. For
{\it weak} attractive interactions, the contribution of the PE
solution to the membrane curvature moduli is small (in units of
$k_BT$), and for repulsive interactions it increases back, and may
reach values of $0.1-1\,k_BT$.

Our work deals only with uniformly charged membranes. In biological
membranes, however, the membrane is composed of a mixture of neutral
and charged lipid. The lipid molecules can rearrange and can cluster
around oppositely charged PE \cite{tzlil}. This in turn can have a
strong effect on the overall membrane rigidity. Future works may
offer extensions of the present one by calculating the contributions
to the spontaneous radius of curvature $R_0$, especially in the case
of asymmetrical solutions, or when the membranes are composed of
asymmetric inner and outer leaflets. Other potential directions may
include changes in the effective lipid headgroup size and water
activity due to the presence of polyelectrolytes.

\vskip 3truecm

{\it Acknowledgments}: The authors wish to thank Michael Kozlov for
helpful discussions. Support from the Israel Science Foundation
(ISF) under grant no. 160/05 and the US-Israel Binational
Foundation (BSF) under grant no. 287/02 is gratefully
acknowledged.

\vspace{1.5cm}
\noindent \textbf{Adi Shafir,$^a$ David Andelman$^a$}\\
\noindent $^a$ \textsl{School of Physics and Astronomy   \\
    Raymond and Beverly Sackler Faculty of Exact Sciences \\
    Tel Aviv University, Ramat Aviv, Tel Aviv 69978, Israel.\\}


\clearpage \onecolumn

\begin{figure}[!ht]
\begin{center}
\centerline{\includegraphics[keepaspectratio=true,width=190mm,clip=true]{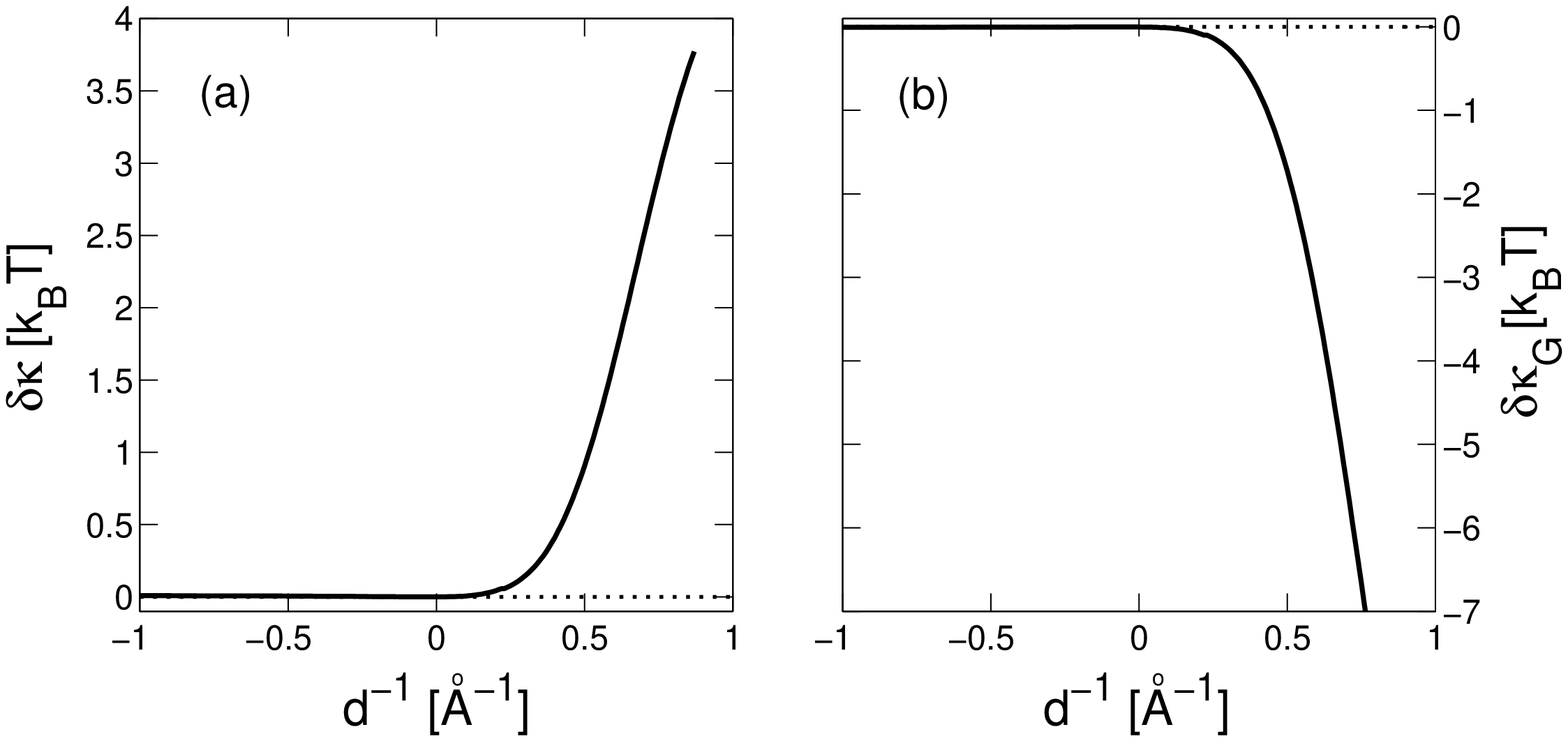}}
\caption{ The $d^{-1}$ dependence of (a) $\delta\kappa$ and (b)
    $\delta\kappa_G$
    is presented for low added salt concentration ($\cs=0.1$\,mM).
    Other parameters are $\scharge=0.001$\,\AA$^{-2}$, $a=5$\,\AA, $f=0.5$,
    $v=50$\,\AA$^3$,
    $\phi_b^2=10^{-8}$\,\AA$^{-3}$,  $T=300$\,K and $\varepsilon=80$.
    For large $d^{-1}$, we see a
    significant increase in the magnitude of both $\delta\kappa$ and $\delta\kappa_G$,
    amounting to several $k_BT$, which is very
    significant in comparison to normal membrane curvature moduli.}
\label{fig3}
\end{center}
\end{figure}

\begin{figure}[!ht]
\begin{center}
\includegraphics[keepaspectratio=true,width=175mm,clip=true]{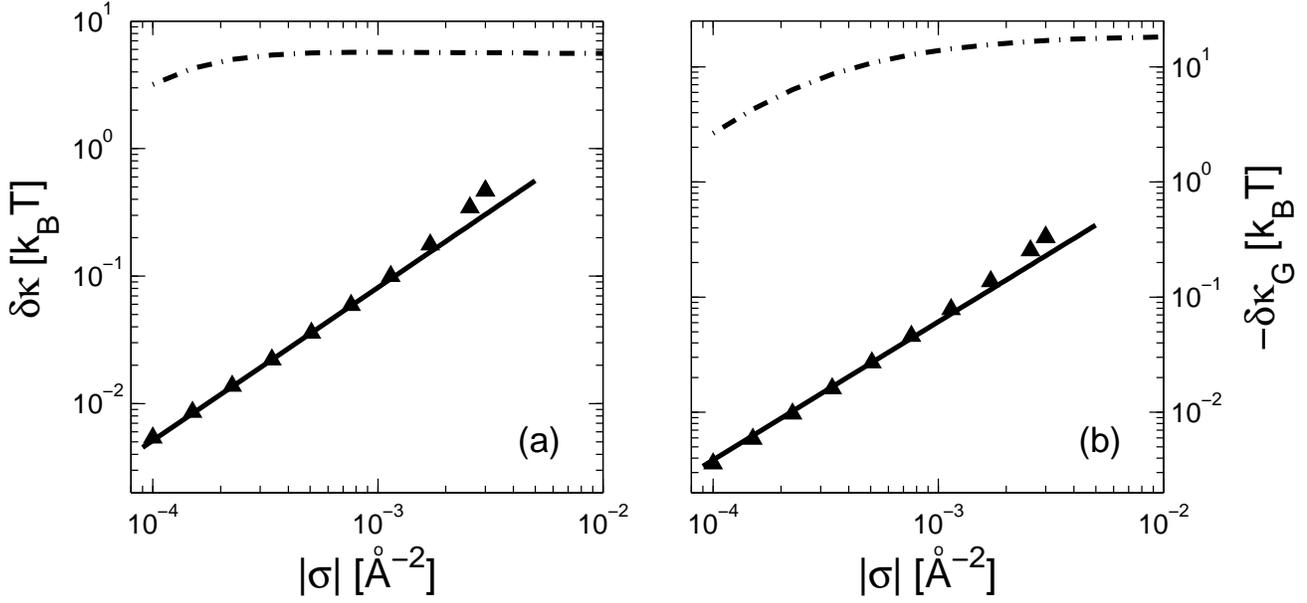}
\caption{    The dependence of $\delta\kappa$ (a) and
$\delta\kappa_G$ (b) on the surface charge
    is presented for the case of repulsive membranes. The
    triangular  symbols are the numerically calculated $\delta\kappa$
    for $\cs=0.1$\,mM, $a=10$\,\AA, $f=0.5$, $v=50$\,\AA$^3$,
    $\phi_b^2=10^{-8}$\,\AA$^{-3}$,  $T=300$\,K and $\varepsilon=80$.
    The solid line scales as $\scharge^{\beta}$ with $\beta\simeq 1.2$.
    As can be seen, for low salt concentrations the  exponent of the curvature
    modulus, $\beta\simeq 1.2$, is close to the predicted $\beta=1$
    derived
    from Eq.~(\ref{kScRSLS}). The dashed-dotted line is the numerically
    calculated $\delta\kappa,\,\delta\kappa_G$ for the case of an
    ionic solution with no polymers, with parameters
    $\cs=0.1$\,
    mM, $T=300$\,K and $\varepsilon=80$.
    The addition of polyelectrolytes can, in this
    case, reduce the curvature moduli significantly.}
\label{fig5}
\end{center}
\end{figure}

\begin{figure}[!ht]
\begin{center}
\centerline{\includegraphics[keepaspectratio=true,width=190mm,clip=true]{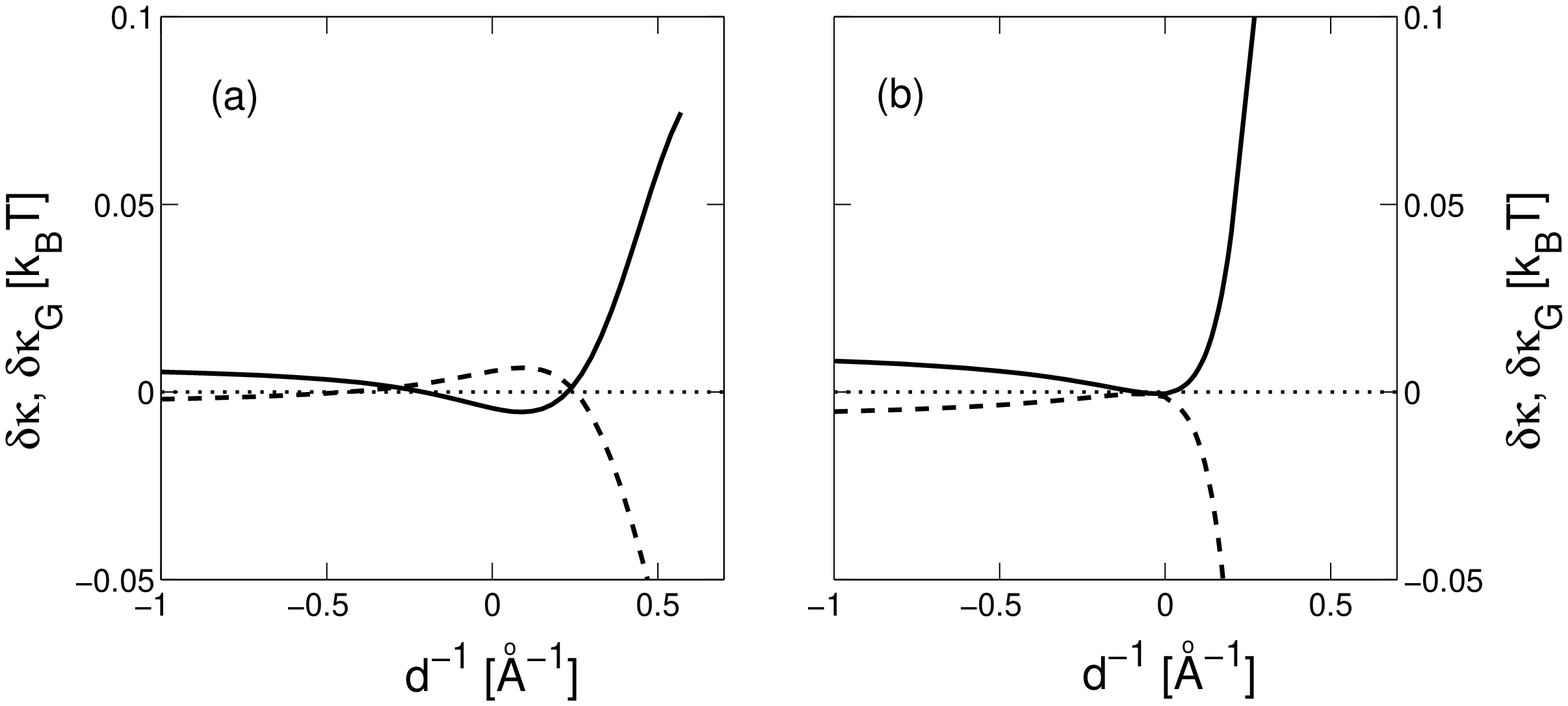}}
\caption{ The $d^{-1}$ dependence of $\delta\kappa$ and
$\delta\kappa_G$ is
    presented for two salt concentrations: (a) $\cs=0.1$\,M and
    (b) $\cs=0.1$\,mM.
    Other parameters are $\scharge=0.001$\,\AA$^{-2}$, $a=5$\,\AA, $f=0.5$, $v=50$\,\AA$^3$,
    $\phi_b^2=10^{-8}$\,\AA$^{-3}$,  $T=300$\,K and $\varepsilon=80$.
    In both plots, the solid line corresponds to $\delta\kappa$ and the dashed
    one to $\delta\kappa_G$. Three regimes for both curvature
    moduli are seen. For $d^{-1}<0$ we obtain $\delta\kappa>0$ and $\delta\kappa_G<0$, for
    $d^{-1}\sim0$ the sign of both is inverted, and for $d^{-1}\gg 0$
    both moduli return to their original sign. The magnitude of both
    moduli is very small for this parameter range.}
\label{fig1}
\end{center}
\end{figure}


\begin{thebibliography}{99}



\bibitem{itamar1} I. Borukhov, D. Andelman, H. Orland,
    {\it Macromolecules}, 1998,
    {\bf 31}, 1665; {\it Europhys. Lett.}, 1995, {\bf 32}, 499.

\bibitem{itamar2} I. Borukhov, D. Andelman, H. Orland, {\it Eur. Phys. J.
    B}, 1998, {\bf 5}, 869.

\bibitem{itamar3} I. Borukhov, D. Andelman, H. Orland, {\it J. Phys. Chem.
    B}, 1999, {\bf 103}, 5042.

\bibitem{us1} A. Shafir, D. Andelman, R.R. Netz,
    {\it J. Chem. Phys.}, 2003, {\bf 119}, 2355.

\bibitem{us2} A. Shafir, D. Andelman, {\it Phys. Rev. E},
    2004, {\bf 70}, 061804.

\bibitem{wang1}  Q. Wang, {\it Macromolecules}, 2005, {\bf 38}, 8911.

\bibitem{wiegel} F.W. Wiegel, {\it J. Phys A: Math. Gen.}, 1977, {\bf 10},
     299.

\bibitem{muthu} M. Muthukumar, {\it J. Chem. Phys.}, 1987, {\bf 86}, 7230.

\bibitem{joanny} J.F. Joanny, {\it Eur. Phys. J. B.}, 1999, {\bf 9}, 117.

\bibitem{varoqui1} R. Varoqui,{\it J. Phys. II (France)}, 1993, {\bf 3}, 1097.

\bibitem{varoqui2} R. Varoqui, A. Johner, A. Elaissari, {\it J. Chem.
    Phys.},  1991, {\bf 94}, 6873.

\bibitem{borisov} O.V. Borisov, E.B. Zhulina, T.M. Birshtein,
     {\it J. Phys. II (France)}, 1994, {\bf 4}, 913.

\bibitem{dobrynin} A.V. Dobrynin, A. Deshkovski, M. Rubinstein,
    {\it Macromolecules}, 2001, {\bf 34}, 3421.

\bibitem{usphase} A. Shafir, D. Andelman, {\it Phys. Rev. E.}, 2006, {\bf 74},
    021803.

\bibitem{decher_book} G. Decher, J. B. Schlenoff, {\it Multilayer Thin
    Films}, Wiley-VCH, Weinheim (2002).

\bibitem{bensimon} D. Bensimon, F. David, S. Leibler, A. Pumir, {\it J.
    Phys. (France)}, 1990, {\bf 51}, 689.

\bibitem{fodnin} A. Fodgen, B. W. Ninham, {\it Langmuir}, 1991, {\bf 7}, 590.

\bibitem{winter} M. Winterhalter, W. Helfrich, {\it J. Phys.
    Chem.}, 1988, {\bf 92}, 6865.

\bibitem{winter2}  M. Winterhalter, W. Helfrich, {\it J. Phys.
    Chem.}, 1992, {\bf 96}, 327.


\bibitem{KK} M. Kiometzis, H. Kleinert, {\it Phys. Lett. A}, 1989,
     {\bf 140}, 520.

\bibitem{lekker} H.N.W. Lekkerkerker, {\it Physica A}, 1989, {\bf 159}, 319.

\bibitem{mitchnin} D. J. Mitchell, B. W. Ninham, {\it Langmuir}, 1989,
    {\bf 5}, 1121.

\bibitem{fodmitchnin} A. Fodgen, D. J. Mitchell, B. W. Ninham,
    {\it Langmuir}, 1990, {\bf 6}, 159.

\bibitem{andelman} D. Andelman,  in {\it  Handbook of Biological Physics:
                 Structure and Dynamics of Membranes},  Vol. 1B,
                 ed. R. Lipowsky and E. Sackmann,
                 Elsevier Science B.V., Amsterdam, 1995, Chap. 12, p. 603.

\bibitem{deVries} H. von Berlepsch, R. de Vries, {\it Eur. Phys. J. E}, 2000,
 {\bf 1}, 141.

\bibitem{scheutjens} P. A. Barneveld, D. E. Hesselink, F. A. M. Leermakeers,
    J. Lyklema, J. M. H. M. Scheutjens, {\it Langmuir}, 1994, {\bf 10}, 1084.

\bibitem{cstuart} M. M. A. E. Claessens, B. F. van Oort, F. A. M. Leermakers,
    F. A. Hoekstra, M. A. Cohen Stuart, {\it Biophys. J.}, 2004, {\bf 87},
   3882.

\bibitem{degennes91} P. G. de Gennes, {\it J. Phys. Chem}, 1990, {\bf 94},
    8407.

\bibitem{marques} J. Brooks, C. Marques, M. Cates, {\it Europhys.
    Lett.}, 1991,
    {\bf 14}, 713; {\it J. Phys. II (France)}, 1991, {\bf 1},673.

\bibitem{joannycle} F. Clement, J-F. Joanny, {\it J. Phys. II
    (France)}, 1997, {\bf 7}, 973.

\bibitem{skau} K.I. Skau, E.M. Blokhuis, {\it Eur. Phys. J. E}, 2002, {\bf 7},13.

\bibitem{chlipo} C. Heidgarst, R. Lipowsky, {\it J. Phys II (France)}, 1996, {\bf 6},
    1465.

\bibitem{sunglee} W. Sung, S. Lee, {\it Europhys. Lett.}, 2004, {\bf 68}, 596.

\bibitem{netzlipo} M. Breidenich, R. R. Netz, R. Lipowsky, {\it Eur. Phys. J.
    E}, 2001, {\bf 5}, 403.

\bibitem{ligoure1} G. Bouglet, C. Ligoure, A.M. Bellocq, E. Dufourc, G. Mosser,
    {\it Phys. Rev. E.}, 1998, {\bf 57}, 834.

\bibitem{ligoure2} J. Appell, C. Ligoure, G. Porte, {\it J. Stat. Mech.:
    Theory and Exp.}, 2004, P08002.

\bibitem{gommper} G. Gompper, H. Endo, M. Mihailescu, J. Allgaier,
    M. Monkenbusch, D. Richter, B. Jakobs, T. Sottmann, R. Strey
    {\it Europhys. Lett.}, 2001, {\bf 56}, 683.

\bibitem{maugey} M. Maugey, A.M. Bellocq, {\it Langmuir}, 2001, {\bf 17}, 6740.

\bibitem{RouxSafinya}D. Roux, C.R. Safinya, {\it J. Phys. (France)}, 1988, {\bf 49}, 307.

\bibitem{harries1} S. May, D. Harries, A. Ben-Shaul, {\it Biophys. J.}, 2000, {\bf 78}, 1681.

\bibitem{harries2} D. Harries, S. May, W.M. Gelbart, A. Ben-Shaul, {\it Biophys
J.}, 1998, {\bf 75}, 159.

\bibitem{degennes} P.G. de Gennes, {\it Macromolecules}, 1981, {\bf 14}, 1637.

\bibitem{nr} W.H. Press, B.P. Flannery, S.A. Teukolsky, W.T. Vetterling,
         {\it Numerical Recipes in C: The Art of
    Scientific Computing}, Cambridge University, Cambridge, 1992, Chap. 17, p. 762.

\bibitem{helfrich} W. Helfrich, {\it Z. Naturaorsch.}, 1973, {\bf 28c}, 693.

\bibitem{skau2} K.I. Skau, E.M. Blokhuis, {\it Macromolecules}, 2003,
    {\bf 36},4637.

\bibitem{frantescu} A. Frantescu, S. Kakorin, K. Toensing, E. Neumann,
    {\it Phys. Chem. Chem. Phys.}, 2005, {\bf 7}, 4126.

\bibitem{tzlil} S. Tzlil and A. Ben-Shaul, {\it Biophys. J.}, 2005,
{\bf 89}, 2972.


\end{thebibliography}
\end{document}